# Physical Action Categorization using Signal Analysis and Machine Learning

Asad Mansoor Khan[1], Ayesha Sadiq[1], Sajid Gul Khawaja[1], Norah Saleh Alghamdi[2], Muhammad Usman Akram[1], Ali Saeed[3]

1. *College of E&ME, National University of Sciences and Technology, Islamabad, Pakistan*
2. *Department of Computer Sciences, College of Computer and Information Sciences, Princess Nourah bint Abdulrahman University, P.O. Box 84428, Riyadh 11671, Saudia Arabia*
3. *National University of Modern Sciences, Pakistan*

**Abstract**

**Daily life of thousands of individuals around the globe suffers due to physical or mental disability related to limb movement. The quality of life for such individuals can be made better by use of assistive applications and systems. In such scenario, mapping of physical actions from movement to a computer aided application can lead the way for solution. Surface Electromyography (sEMG) presents a non-invasive mechanism through which we can translate the physical movement to signals for classification and use in applications. In this paper, we propose a machine learning based framework for classification of 4 physical actions. The framework looks into the various features from different modalities which contribution from time domain, frequency domain, higher order statistics and inter channel statistics. Next, we conducted a comparative analysis of k-NN, SVM and ELM classifier using the feature set. Effect of different combinations of feature set has also been recorded. Finally, the classifier accuracy with SVM and 1-NN based classifier for a subset of features gives an accuracy of 95.21 and 95.83 respectively. Additionally, we have also proposed that dimensionality reduction by use of PCA leads to only a minor drop of less than 5.55% in accuracy while using only 9.22% of the original feature set. These finding are useful for algorithm designer to choose the best approach keeping in mind the resources available for execution of algorithm.**

## 1. Introduction

In the modern day, physical disabilities present a major problem to the daily life. The main reason behind this is the various factors that contribute to these disabilities. These factors include gait disorder or limb impairment due to aging process [1], occupational injuries or trauma such as sports accidents, which hinder the quality of life. Stroke is another major cause of limb disabilities in adults [2]. Most of these sufferers may require partial limb support or prosthetics limb to elevate their daily suffering. Apart from these factors, neurological disorders are also a leading cause of accidents [3]. Epilepsy, a major neurological disorder, is caused by unusual nerve cell activity in the brain [4], which effects almost 50 million of people around the globe [5]. This brings to light the immense need for a system that can categorize the physical signals for either prosthetic limb design or timely notification of epileptic attacks for injury prevention.

In this regard, a possible solution is to somehow sense the intended motion and take decisions accordingly. Surface Electromyography (sEMG) has been characterized as the best non-invasive performer for activity analysis [6]. Electromyography (EMG) refers to the electrical activity recordings, which are produced because of skeletal muscles. Fig 1. shows a side by side comparison of sEMG signals collected against normal and abnormal activities such as clapping and elbowing respectively. These EMG can be used to analyze the biomechanics of human or animal movement. Due to this, sEMG is used both in the identification of ailments of the muscular system, clinical or biomedical applications and in the development of modern human computer interaction [3]. These signals can be analyzed to detect medical abnormalities [7, 8], emotion detection [9], prosthetic arms, hand and lower limbs control [10]. Various approaches have been suggested by different people to analyze EMG signal.

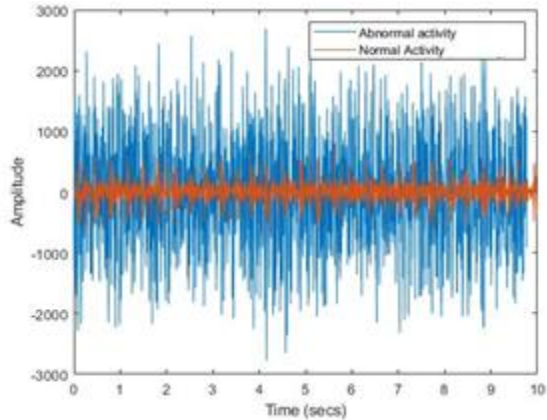

Fig 1. Raw sEMG signal for normal and abnormal activities signifying the difference

A new classification algorithm for multiple physical actions using sEMG is proposed in this paper. In the proposed approach, a window of raw sEMG signal is initially pre-processed to enhance the variability between signals of normal and abnormal activates. The processed window is then forwarded to feature extractor where, among the well-known features for EMG, inter channel correlation and frequency based signatures for classes is also calculated. The feature vector is then subjected to normalization and is finally fed to a classifier to provide us with the output label of the signal. The proposed methodology results in high classification accuracy even with use of a simple classifier and lower number of features as compared to previous methods.

The rest of the paper is structured in the following manner: Section 2 discusses state-of-art algorithms proposed by researchers for detection of physical activity using EMG signals. Proposed methodology is presented in Section 3, which is followed by dataset explanation, experimentation, and results, which are under Section 4. Finally, Discussion on these results and conclusion are stated in Sections 5 and 6 respectively.

## 2. Literature Review

Surface electromyography is a discipline that studies or identifies electrical activity of muscles to recognize morphological variations of the neuromuscular system. Electromyography (sEMG) signal is an electrical activity produced when nervous and muscular events are recorded from the skin surface of human skeletons using electrodes, which can reflect the functionality of nerves and muscles on a real-time basis. Nevertheless, how to proficiently exact features from electromyography signals to understand accurate identification of events is the main issue to attain precision of rehabilitation cure and preparation of electromyography-controlled prostheses.

An extreme learning machine algorithm based on EMG signal using bispectrum and QPCs (quadratic phase coupling) for all the dataset in order to classify aggressive and normal activity is proposed in [11]. Improved EMD method based on median filtering and feature extraction for the classification of ALS (amyotrophic lateral sclerosis and Normal EMG signals demonstrated in [12]. A novel approach of using time-frequency representation of an EMG signal and convolutional neural network in order to distinguish between normal and aggressive action is suggested in [13]. Discrimination of Aggressive actions from normal actions based on adaptive neuro-fuzzy inference system (ANFIS) and feature extraction from EMG signals is described in [14]. In this study, eight channels recorded the EMG signals of 10 aggressive and 10 normal actions of three males and one female, which were then analyzed for classifying normal and aggressive actions. An improved classification framework depends upon modified spectral moment based features and inter-channel correlation feature proposed in [15]. Identification of physical activities of sEMG signal based on variational mode decomposition (VMD), statistical feature extraction and multi class least square support vector machine are discussed in [16]. Classification of physical actions as normal and aggressive using quadratic phase coupling and artificial neural network is described in [17]. An EMG pattern recognition system in which multi-scale principal component analysis is used for de-noising whereas discrete wavelet transform based feature extraction is proposed in [18]. Another approach in which time-frequency Image is used as input to pre-trained convolutional neural networks. Deep feature extraction using AlexNet and VGG-16 and SVM is used for the classification of EMG based Physical activities in [19].

An extreme learning machine (ELM) classifier based on flexible analytic wavelet transform features for the identification of physical actions is proposed in [20]. A non-invasive technique to provide reference control signals for prosthetic hands based on SEMG signal carried out feature extraction using wavelet analysis and contributes information in time-frequency domain whereas classification system was done using artificial neural networks (ANNs) provides in [21]. Evaluation SEMG quality by performing comparison of five different classifiers, two of them are supervised whereas three of them are unsupervised artificial neural networks (ANNs). Unsupervised artificial neural networks perform better classification accuracy which is greater than 98% as compared to supervised artificial neural networks describes in [22]. Use of convolutional neural networks (CNNs) for feature extraction of SEMG signals and perform classification of actions. Because of local connection and weight sharing properties of CNNs, it exhibit good translation invariance. Consequently, the spectrogram acquired by evaluating SEMG signals and used as an image to CNN is demonstrated in [23].

## 3. Methodology

In our paper, we address the problem of M-class classification of physical actions which is based on C-channel sEMG data. The proposed methodology, in this regard, is divided into pre-processing of RAW sEMG signals, feature extraction and selection, feature normalization and classification into M-classes. The system level flow diagram representing the proposed methodology is presented in Fig 2.

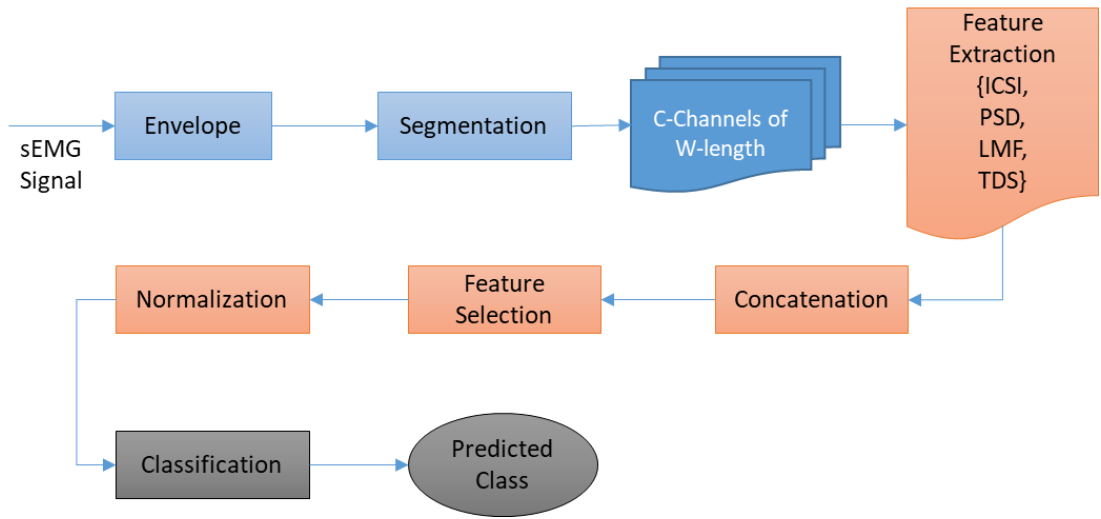

Fig 2. System level flow diagram of our proposed methodology

*3.1. Pre-Processing and Segmentation*

The first step in our proposed methodology is to pre-process and segment out each channel of sEMG signal. In order to enhance the interclass variation between aggressive and normal actions upper envelope of sEMG signal is calculated using Hilbert Transform. Qualitatively, envelope of any signal represents the upper and lower boundary limits within which signal is contained.

The resultant signal after envelop calculation is subjected to segmentation where W-length windows with an overlap factor of 25% are generated. The length "W" of the window is controlled by the sampling frequency through which the concerned signal is acquired. Note that after segmentation now each C-channel sEMG signal is converted to Ns samples of length W having C-channels. Now feature extraction is performed on each of the sub-windows from every pattern.

## 3.2. Feature Vector Generation

In order to generate a valid feature vector our proposed methodology makes use of diverse features from time and frequency domain thus motivating us to employ feature selection to trim the number of features. Moreover, feature normalization is conducted on the selected features so that contribution of each feature becomes approximately proportionately to the final distance while classifying.

### 3.2.1. Feature Extraction
The feature vector for our proposed methodology contains signatures from various modalities including Time Domain, Frequency Domain and Inter-Channel Statistics. These features are elaborated in the relevant subsections.

### 3.2.1.1. Inter Channel Statistics
The first subgroup of our feature vector consists of statistics that are calculated between the corresponding $w^{th}$ segments of all $M$ channels. The channel wise paring for the channels is shown in Tab 1. In all a total of 56 features belong to this subgroup of features whose division is discussed below.

Tab 1. Channel wise pairing for inter-channel statistic calculation

| Features | Channel Pair | Features | Channel Pair | Features | Channel Pair | Features | Channel Pair |
|---|---|---|---|---|---|---|---|
| 1 | (1, 2) | 8  | (2, 3) | 15 | (3, 5) | 22 | (4, 8) |
| 2 | (1, 3) | 9  | (2, 4) | 16 | (3, 6) | 23 | (5, 6) |
| 3 | (1, 4) | 10 | (2, 5) | 17 | (3, 7) | 24 | (5, 7) |
| 4 | (1, 5) | 11 | (2, 6) | 18 | (3, 8) | 25 | (5, 8) |
| 5 | (1, 6) | 12 | (2, 7) | 19 | (4, 5) | 26 | (6, 7) |
| 6 | (1, 7) | 13 | (2, 8) | 20 | (4, 6) | 27 | (6, 8) |
| 7 | (1, 8) | 14 | (3, 4) | 21 | (4, 7) | 28 | (7, 8) |

I. *Maximum Similarity Index*
   *These* statistics are based on maximum cross correlation [24] among the corresponding segments of two channels a and b of an EMG signal which is defined as
   $$C_{i,j} = \max\left(CrossCorrelation(seg_i, seg_j)\right) \text{ Where } i \,!= j$$
   The above equation represents the maximum correlation between the segments of two channels $s_a$ and $s_b$ of an EMG signal. As we have 8 channels so we can get 28 values after performing the maximum correlation between corresponding channels.

II. *Covariance Index*
   These statistics belong to the class of higher order statistics, which represent the cross-correlation between respective segments of all channel pairs.

### 3.2.1.2. Power Spectral Density
The spectral features were previously proposed for the identification of an EMG pattern in [25]. In the proposed technique, the spectral band power features are extracted using Burgs Transform. For each channel of an EMG pattern, assuming a model order $v$, the power spectral density is estimated as:

$$\psi(w_k) = |H(w_k)|^2 \sigma^2_{burg}$$

Here $\sigma^2_{burg}$ is the error variance computed in the Burg's method. Finally, the PSD features are evaluated by dividing the spectrum into $N_b$ bands and computing the respective powers in those bands as:

$$\eta_b^m = \sum_{k \in b^{th} band} \psi(w_k)$$

Thus a total of $N_b$ features against each segment of every channel are calculated leading to a feature vector of $8N_b$.

### 3.2.1.3. Log Moments of Fourier Spectra

The logarithms of moments and their ratios from the frequency domain are computed for the EMG segments based on [27]. The $i^{th}$ frequency domain moment from the spectrum is defined as [28] where $\psi(w_k)$ represents the magnitude response against $w^{th}$ segment of $m^{th}$ channel

$$g_j(i) = \sqrt{\sum_{k=1}^{L} k^i \psi_j(k)}$$

Based on the above expression a total of 17 features have been calculated using the given expressions

$$f_j(1) = \ln g_j(0)$$
$$f_j(2) = \ln g_j(2)$$
$$f_j(3) = \ln g_j(4)$$
$$f_j(4) = \ln g_j(0) - 1/2\ln(g_j(0) - g_j(2)) - 1/2\ln(g_j(0) - g_j(4))$$
$$f_j(5) = \ln g_j(2) - 1/2\ln(g_j(0)g_j(4))$$
$$f_j(6) = \ln g_j(0) - 1/2\ln(g_j(1)g_j(3))$$
$$f_j(7) = \ln g_j(0) - 1/2\ln(g_j(2)g_j(6))$$

The pair wise features based on moment products are

$$f_j(n) = 1/2\ln(g_j(i)g_j(j))$$

With $n = 8, \ldots, 17$ where the values of i and j chosen from the set $C_2$ of 10 ordered pairs defined below

$$C_2 = \{(i, j) : i = 1,2,3,4; j = 2,3,4,5 \text{ and } i \neq j\}$$

### 3.2.1.4. Time Domain/Miscellaneous Features

This subset of features correspond to the most common type of time domain-features which are frequently used for sEMG signal analysis [29, 30, 31, and 32]. These features are listed in Tab

- *Amplitude:* The maximum amplitude of the signal
- *Root Mean Square:* The RMS represents the square root of the average power of the EMG signal for a given period of time.
- *Variance:* Variance of EMG signal (VAR) is good at measuring the signal power
- *Waveform length:* WL is a cumulative variation of the EMG that can indicate the degree of variation about the EMG signal.
- *Mean Absolute Value:* The Mean absolute value (MAV) is one of the most popular EMG features, and it is defined as the average of the summation of absolute value of signal.
- *Simple Square Integral:* Simple square integral (SSI) is defined as the summation of square values of EMG signal amplitude
- *Zero Crossing:* ZC counts the times that the signal changes sign.
- *Slope Sign Change:* SSC counts the times the slope of the signal changes sign.
- *Willison Amplitude:* WAMP is the number of counts for each change of the EMG signal amplitude between two adjacent samples that exceeds a defined threshold.
- *Integrated EMG:* Integrated EMG (IEMG) is generally used as a pre-activation index for muscle activity. It is the area under the curve of the rectified EMG signal. IEMG can be simplified and expressed as the summation of the absolute values of the EMG amplitude.
- *Log detector:* It is a feature that is good at estimating the exerted force.
- *Myopulse percentage rate:* MYOP is defined as the mean of Myopulse output in which the absolute value of EMG signal exceeds the pre-defined threshold value.
- *Difference absolute standard deviation value:* DASDV is another frequently used EMG feature

- *Enhanced Mean Absolute value:* It is a feature that is good at estimating the exerted force
- *Enhanced Wavelength:*
- *Modified Mean Absolute Value:* Modified mean absolute value (MMAV) is an extension of MAV feature by assigning the weight window function.
- *Modified Mean Absolute Value 2:* Modified mean absolute value 2 (MMAV2) is another extension of MAV feature by assigning the continuous weight window function.
- *Maximum Fractal Length:* MFL is a recently established method for measuring low-level muscle activation. When the smallest scale is set to one, the definition of MFL resembles a modified version of WL by using the RMS and logarithm functions.
- *Average Amplitude Change:* Average amplitude change (AAC) is nearly equivalent to WL feature, except that wavelength is averaged.
- *Kurtosis:* Kurtosis is known as a statistical method that used to describe the distribution and a characteristic that identifies the tendency of peak data. Kurtosis level data is determine by comparing the peak of the curve inclination data.
- *Skewness:* Skewness is defined as the inclination distribution data.

### 3.2.1.5. Higher Order Statistics

This subgroup belongs to two higher order statistics based features, which are calculated against every $w^{th}$ segments of all M channels. Firstly second order Cumulant, representing alternatives to moment distribution, is calculated against each channel leading to eight features. Secondly, fourth order Cross-Cumulants is calculated between the channels leading to two features against upper limb and lower limb respectively.

The calculation of all the above features leads to a feature vector of length 303, which is summarized in Tab 2.

Tab 2. Summary of calculated and selected features

| Feature Type | Feature Count |
|---|---|
| ICS | 56 |
| PSD | 80 |
| LMFS | 136 |
| TDS | 21 |
| HOSA | 10 |
| **Total** | **303** |

### 3.2.2. Feature Normalization

After calculation and selection of features, the resultant feature vector is normalized before proceeding to classification. Z-score normalization has been employed, using equation 2, on the feature vector so that the value for all features comes within range of [1, -1].

$$x' = \frac{(x - \mu)}{\sigma}$$

Here $x$ represents the feature value, $\sigma$ and $\mu$ represent standard deviation and mean values for feature $x$.

### 3.3. Classifier

The extracted feature vector mentioned in section 3.2 is passed onto a classifier to predict the action class out of 4 possible actions. In our work, we have implemented various classification schemes with focus towards using a simple classifier. The implemented and tested classifiers include K-Nearest Neighbor, Support Vector Machine and Extreme Learning Machines. The proposed methodology explained in this section is implemented on the sEMG data which has been collected locally and obtained results are discussed extensively for the performance evaluation of methodology.

## 4. Experimentation and Results

### 4.1. Dataset

Physical action dataset for the initial analysis of our proposed methodology is self-generated by the use of Thalmic Myoware Armband. The dataset contains sEMG data of one male subject performing four actions, the details of these actions are mentioned in Tab 1. sEMG collected from these subjects has 8 channels, where each channel corresponds to time series data from each electrode which consists of approximately 10000 values.

Tab 2. Summary of physical actions

| Actions |
|---|
| Typing |
| Rest |
| Lifting |
| Pushups |

### 4.2. Experimentation

The physical activity dataset mentioned previously contains approximately 10000 samples per action per subject. In the first step these samples are subdivided into multiple overlapping segments where length of each segment is set to 1000 with an overlap of 25%. This division of each sample against each subject and action leads to a sample space of 974 sample where samples per action are between 45-52 depending upon the length of original recording. This sample space is used to extract a feature vector against each segment using different modalities which have been discussed in section 3. The experimental setup for validation consists of measuring this parameter for a 4-class problem. SVM, K-NN and ELM have been used to classify using complete set of features and their subset. The feature set has been subdivided into subsets containing individual feature type and its possible combinations to get the best classification rate. In view of measuring the performance of our proposed algorithm, different parameters including Accuracy, Sensitivity, Specificity, F-measure and precision have been calculated. 10 fold cross validation for SVM and K-NN is used for classification of our data into 20 and 10 class each. Initially, the complete feature vector of length 303 is passed to the classifiers to classify data into 20, and 10 class respectively for normal and aggressive actions.

Fig 3. represents the classification accuracies using various feature subsets using SVM and 1-NN.

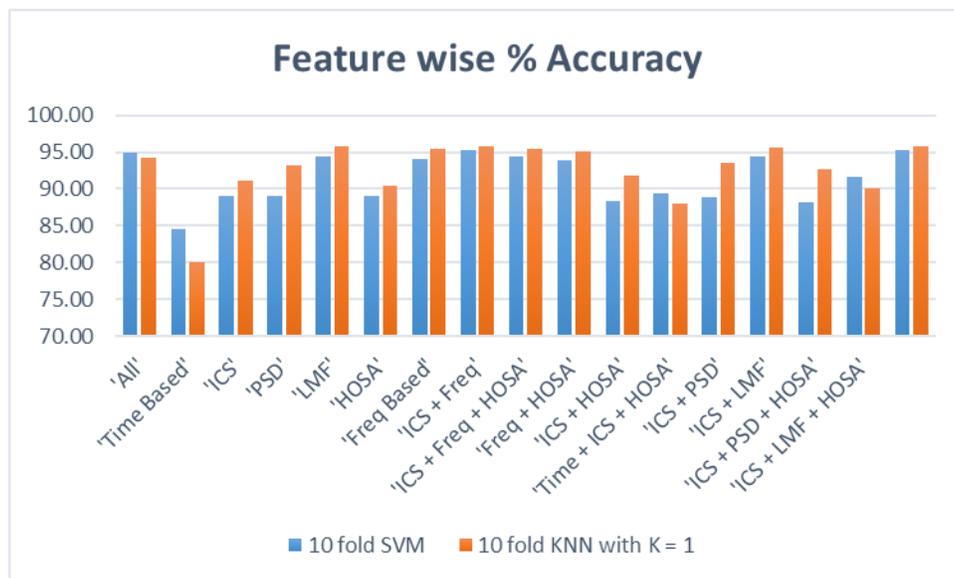

Fig 3. Feature wise Accuracies using SVM and 1-NN classification

Tab 3 shows the classification accuracies using K-NN classifier for K = {1, 2, 3,….,10} for 10 tries and 10 fold cross-validation to avoid biasness in result. The best results are achieved for K = 1 and feature subset of Inter channel statistics and frequency based features.

Tab 3. Classification Summary using 10-fold SVM and selected features

| Features | All Actions | Normal Actions | Aggressive Actions |
|---|---|---|---|
| 'All' | 90.25 | 82.36 | 98.22 |
| 'Time Based' | 52.81 | 27.86 | 77.31 |
| 'ICS' | 84.99 | 78.66 | 92.59 |
| 'PSD' | 80.41 | 69.21 | 92.71 |
| 'LMF' | 81.56 | 68.88 | 95.65 |
| 'HOSA' | 61.97 | 51.89 | 79.34 |
| 'Freq Based' | 87.70 | 76.21 | 98.18 |
| 'ICS + Freq' | 90.49 | 81.63 | 98.26 |
| 'ICS + Freq + HOSA' | 90.55 | 82.19 | 98.30 |
| 'Freq + HOSA' | 88.36 | 77.80 | 98.30 |
| 'ICS + HOSA' | 86.84 | 79.63 | 92.59 |
| 'Time + ICS + HOSA' | 87.93 | 79.31 | 96.27 |
| 'ICS + PSD' | 87.37 | 79.55 | 94.82 |
| 'ICS + LMF' | 88.11 | 77.35 | 97.64 |
| 'ICS + PSD + HOSA' | 88.44 | 80.53 | 96.19 |
| 'ICS + LMF + HOSA' | 86.10 | 77.52 | 95.28 |

As for the ELM classifier 8 different activation functions are used in the hidden neurons. The number of hidden neurons for our problem has been set to 200, which is selected after experimentation. Data division of 80-20 has been used with 20 tries with shuffled data and its division to avoid any learning bias in the classification. Tab 4. shows the training accuracies against different sigmoid functions.

Tab 4. Average Training Accuracy using ELM Classifier with 200 neurons in the hidden layer

| | 'sig' | 'sin' | 'hardlim' | 'tribas' | 'radbas' | 'relu' | 'lrelu' | 'smax' |
|---|---|---|---|---|---|---|---|---|
| **Training Accuracy** | 1.00 | 1.00 | 1.00 | 1.00 | 1.00 | 1.00 | 1.00 | 0.997 |

ELM shows a substantial increase in the training accuracies for all 3 problems, unfortunately this the effect of overfitting and cramming with regard to the training data. Once the testing accuracies are calculated using ELM a major drop in accuracy is seen for all sigmoid functions. Tab 5. represents the testing accuracies for 20 and 10 class problem. The best results are achieved using sigmoid, ReLu and L-Relu activations functions.

Tab 5. Average Testing Accuracy using ELM Classifier with 200 neurons in the hidden layer

| | 'sig' | 'sin' | 'hardlim' | 'tribas' | 'radbas' | 'relu' | 'lrelu' | 'smax' |
|---|---|---|---|---|---|---|---|---|
| **Testing Accuracy** | 0.80 | 0.70 | 0.69 | 0.71 | 0.71 | 0.77 | 0.76 | 0.56 |

The results show that 1-NN outperforms other classifiers by a fraction. The initial results of accuracy for different subsets of feature show that the Inter channel statistics and frequency domain features provide us with the best accuracy. After the initial experimentation the selected features were further reduced by the use of Principal Component Analysis (PCA) to reduce the work load of our classifier. Fig 4. shows the relationship of accuracy loss with reduction in dimensionality using SVM for 4 class problem. The results indicate that using a mere 9.22% of feature vector i.e. 26 features the accuracy comes out to be 89.93% resulting in less than 5.6% loss. Finally, a comparison of Cohen's Kappa values for 1-NN, SVM based classification against all features, selected features and top 26 selected features after PCA returns 0.944, 0, 921, 0.935, 0.921and 0.866 respectively. This shows that using either of these options marginally effects the decision power of the classifier for our problem.

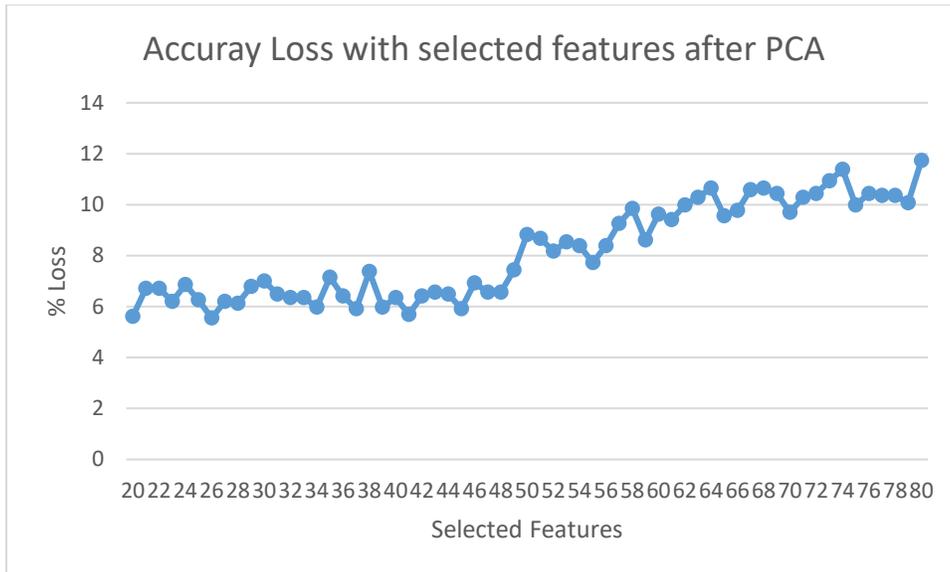

Fig 4. Relationship between accuracy losses against dimensionality reduction using PCA

The overall effect of feature reduction is not significant in classification of physical actions, this can be visualized in Fig 5. where the comparative balanced accuracy against all features, selected features and top 26 PCA based features' which are providing best accuracy as shown in Fig 4.

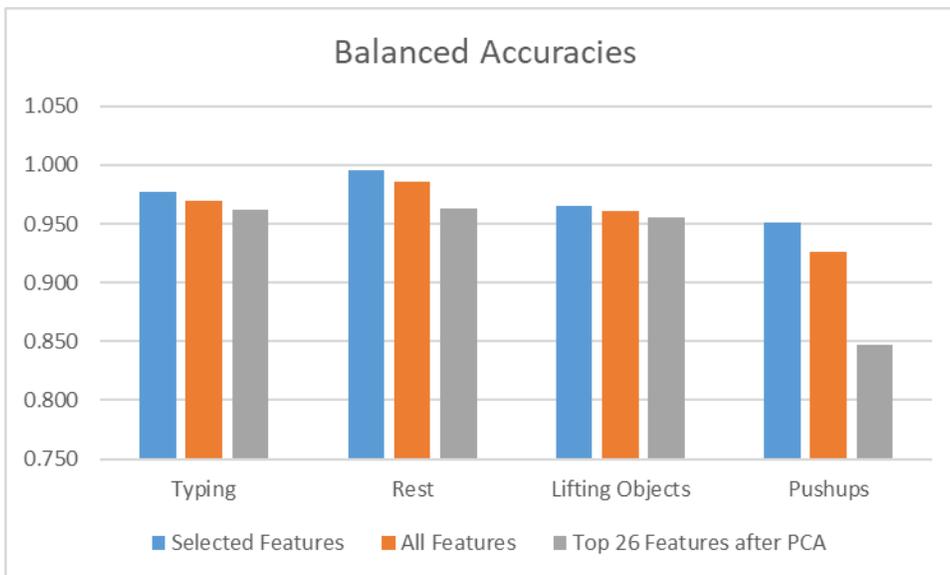

Fig 5. Balanced accuracy for using 1-NN for a) All Features, b) selected feature set (ICS and Frequency) and c) top 26 features using PCA

Moreover, a thorough analysis of SVM for classification of physical actions using complete and subset of features is shown in Tab .7 where comparison between performance measure, such as sensitivity, specificity and precision, is shown. Specificity and sensitivity represent the true negative rate and true positive rate of our classifier against proposed feature set. Precision quantifies the true positive rate that actually belong to that class. The table indicates a marginal change in these values even if a subset of features or even a lower subset of features after dimensionality reduction is used for classification.

Tab 7. Comparison of features based on sensitivity, specificity and precision

| Actions | 1-NN Best Features | 1-NN All Features | SVM Best Features | SVM All Features | Top 26 Features after PCA |
|---|---|---|---|---|---|
| | **Sensitivity** | **Sensitivity** | **Sensitivity** | **Sensitivity** | **Sensitivity** |
| **Typing** | 0.973 | 0.973 | 0.959 | 0.959 | 0.945 |
| **Rest** | 1.000 | 0.986 | 1.000 | 1.000 | 1.000 |
| **Lifting Objects** | 0.944 | 0.931 | 0.958 | 0.931 | 0.917 |
| **Pushups** | 0.917 | 0.875 | 0.889 | 0.875 | 0.736 |
| | **Specificity** | **Specificity** | **Specificity** | **Specificity** | **Specificity** |
| **Typing** | 0.981 | 0.966 | 0.986 | 0.990 | 0.979 |
| **Rest** | 0.990 | 0.985 | 0.985 | 0.985 | 0.926 |
| **Lifting Objects** | 0.986 | 0.990 | 0.986 | 0.981 | 0.995 |
| **Pushups** | 0.986 | 0.977 | 0.977 | 0.963 | 0.958 |
| | **Precision** | **Precision** | **Precision** | **Precision** | **Precision** |
| **Typing** | 0.947 | 0.910 | 0.959 | 0.972 | 0.945 |
| **Rest** | 0.973 | 0.959 | 0.959 | 0.959 | 0.826 |
| **Lifting Objects** | 0.958 | 0.971 | 0.958 | 0.944 | 0.985 |
| **Pushups** | 0.957 | 0.926 | 0.928 | 0.887 | 0.855 |

## 5. Discussion

Non-invasive signal acquisition sensors such as sEMG can play an important role in elevating the life style of people suffering from numerous physical and neurological disabilities/diseases. Correct cataloging of physical actions is the first step in providing a viable solution to such patients. In this article, we have proposed a framework which can help in designing an assistive technologies by providing classification of physical actions.

In our approach, use of pre and post processing greatly affects the classification accuracy, be that for complete feature set or an optimal subset. The results clearly show that SVM and 1-NN using the feature combinations of ICS and Frequency Domain Analysis provide the best classification accuracy in comparison to complete feature set and other classifiers. The respective classification accuracies for 1-NN and SVM are 95.83 and 95.21. Fig. 5 and 6. Clearly indicate that class 4 (pushups) is the one with most inconsistent results among the physical actions.

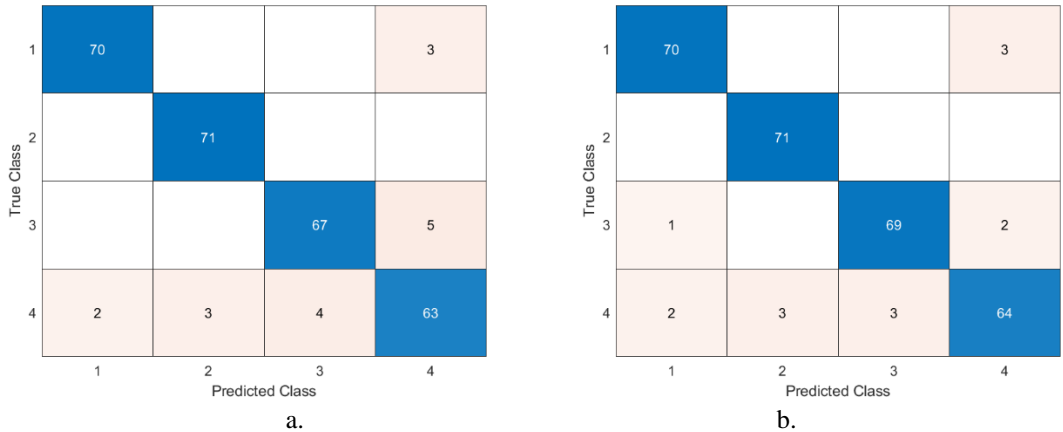

Fig 6. Confusion Matrix f using SVM for a) All Features, b) selected feature set (ICS and Frequency)

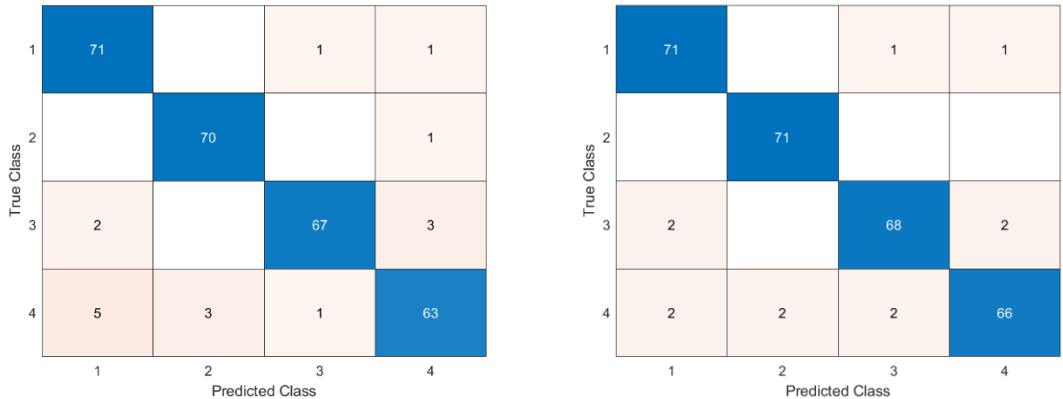

Fig 7. Confusion Matrix using 1-NN for a) All Features, b) selected feature set (ICS and Frequency)

The significance of these classes is also clear if misclassification rate against each class if calculated using SVM. Fig 8. Clearly shows that the maximum misclassification is being caused by pushups.

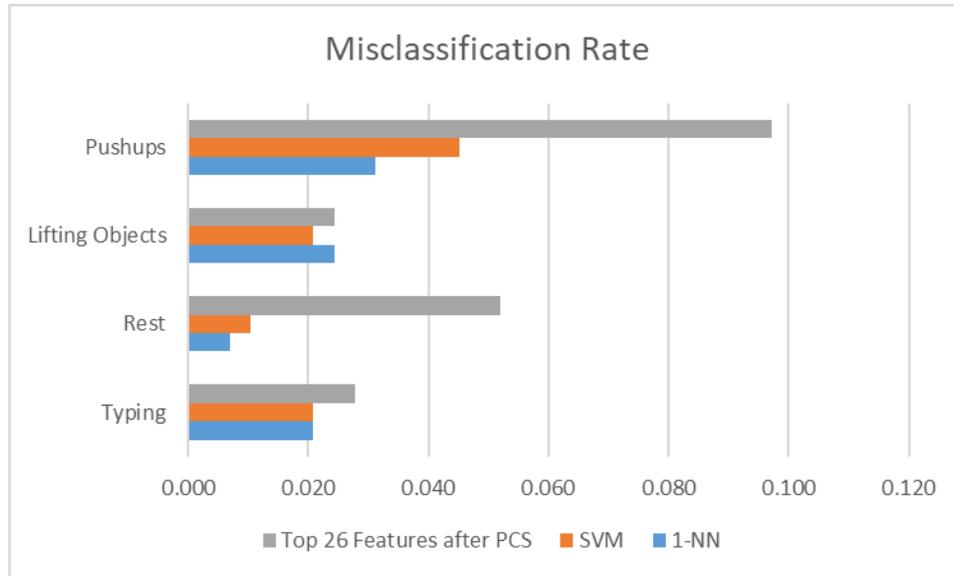

Fig 8. Misclassification rate using SVM for selected feature set (ICS and Frequency) and top 26 features using PCA

The results and the accompanying discussions yield that we can use all three option of feature selection with little loss in accuracy. Though decreasing the number of features can have a positive impact in real-time processing of signals if an embedded solution is sought. As the number of features to the classifier decrease so do the number of tunable parameters thus leading to less number of computation incidentally saving computational resource, time and power.

## 6. Conclusion

In this paper, we have proposed a classification framework for categorizing physical action dataset based on SVM using a feature set calculated from sEMG data. Initially a set of 303 features were calculated, from various domains, 282 were categorized as the meaningful features. 10-fold cross-validation using SVM and 1-NN classification led to an accuracy of 95.21% and 95.83% respectively for the 4 physical actions. The feature vector was further reduced using PCA to indicate that the accuracy falls by a factor of less than 5.6% while using only 9.22% of the original feature vector. These findings are particularly useful for the application designer keeping in view the platform where the algorithm is eventually going to work on. Reducing the feature set can significantly reduce the power and execution time facilitating real-time processing on embedded devices if need be.

**Acknowledgment**
We thank Princess Nourah bint Abdulrahman University Program which provided us with the funding under Researchers Supporting Project Number 'PNURSP2022R40', Princess Nourah bint Abdulrahman University, Riyadh, Saudia Arabia.

**References**

 [1] H. Stolze, S. Klebe, C. Baecker, C. Zechlin, L. Friege, S. Pohle, and G. Deuschl, "Prevalence of gait disorders in hospitalized neurological patients," Movement disorders, vol. 20, no. 1, pp. 89–94, 2005.

[2] T. Proietti, V. Crocher, A. Roby-Brami, and N. Jarrasse, "Upper limb robotic exoskeletons for euro-rehabilitation: A review on control strategies," IEEE Reviews in Biomedical Engineering, vol. 9, pp. 4–14, 2016.

[3] Van Den Broek, Mariska, Ettore Beghi, and RESt-1 Group. "Accidents in patients with epilepsy: types, circumstances, and complications: a European cohort study." Epilepsia 45.6 (2004): 667-672.

[4] Fisher, Robert S., et al. "ILAE official report: a practical clinical definition of epilepsy." Epilepsia 55.4 (2014): 475-482.

[5] https://www.who.int/news-room/fact-sheets/detail/epilepsy

[6] LONG, CHARLES, and MARY ELEANOR BROWN. "Electromyographic kinesiology of the hand: muscles moving the long finger." JBJS 46.8 (1964): 1683-1706.

[7] Bajaj, Varun, and Anil Kumar. "Features based on intrinsic mode functions for classification of EMG signals." International Journal of Biomedical Engineering and Technology 18.2 (2015): 156-167.

[8] Singh, Rajat Emanuel, et al. "A Review of EMG Techniques for Detection of Gait Disorders." Machine Learning in Medicine and Biology. IntechOpen, 2019.

[9]Girardi, Daniela, Filippo Lanubile, and Nicole Novielli. "Emotion detection using noninvasive low cost sensors." 2017 Seventh International Conference on Affective Computing and Intelligent Interaction (ACII). IEEE, 2017.

[10] Tavakoli, Mahmoud, Carlo Benussi, and Joao Luis Lourenco. "Single channel surface EMG control of advanced prosthetic hands: A simple, low cost and efficient approach." Expert Systems with Applications 79 (2017): 322-332.

 [11] Sezgin, Necmettin. "Analysis of EMG signals in aggressive and normal activities by using higher-order spectra." The Scientific World Journal 2012 (2012).

 [12] Mishra, Vipin K., et al. "An efficient method for analysis of EMG signals using improved empirical mode decomposition." AEU-International Journal of Electronics and Communications 72 (2017): 200-209.

[13] Alaskar, Haya Mohammad Abdulaziz. "Deep Learning of EMG Time–Frequency Representations for Identifying Normal and Aggressive Actions." (2018).

[14] Jana, Gopal Chandra, Aleena Swetapadma, and Prasant Pattnaik. "An intelligent method for classification of normal and aggressive actions from electromyography signals." 2017 1st International Conference on Electronics, Materials Engineering and Nano-Technology (IEMENTech). IEEE, 2017.

[15] Turlapaty, Anish C., and Balakrishna Gokaraju. "Feature Analysis for Classification of Physical Actions Using Surface EMG Data." IEEE Sensors Journal 19.24 (2019): 12196-12204.

[16] Sukumar, Nagineni, Sachin Taran, and Varun Bajaj. "Physical actions classification of surface EMG signals using VMD." 2018 International Conference on Communication and Signal Processing (ICCSP). IEEE, 2018.

[17] Šahinbegović, Hana, Laila Mušić, and Berina Alić. "Distinguishing physical actions using an artificial neural network." 2017 XXVI International Conference on Information, Communication and Automation Technologies (ICAT). IEEE, 2017.

[18] Podrug, Ermin, and Abdulhamit Subasi. "Surface EMG pattern recognition by using DWT feature extraction and SVM classifier." The 1st conference of medical and biological engineering in Bosnia and Herzegovina (CMBEBIH 2015). 2015.

[19] Demir, Fatih, et al. "Surface EMG signals and deep transfer learning-based physical action classification." Neural Computing and Applications 31.12 (2019): 8455-8462.

[20] Sravani, C., et al. "Flexible analytic wavelet transform based features for physical action identification using sEMG signals." IRBM 41.1 (2020): 18-22.

[21] Hatem, Samar M., et al. "Rehabilitation of motor function after stroke: a multiple systematic review focused on techniques to stimulate upper extremity recovery." Frontiers in human neuroscience 10 (2016): 442.

[22] Akhundov, Riad, et al. "Development of a deep neural network for automated electromyographic pattern classification." Journal of Experimental Biology 222.5 (2019): jeb198101.

[23] Duan, Na, et al. "Classification of multichannel surface-electromyography signals based on convolutional neural networks." Journal of Industrial Information Integration 15 (2019): 201-206.

[24] Smith, Lauren H., and Levi J. Hargrove. "Comparison of surface and intramuscular EMG pattern recognition for simultaneous wrist/hand motion classification." 2013 35th annual international conference of the IEEE engineering in medicine and biology society (EMBC). IEEE, 2013.

[25] Khushaba, Rami N., et al. "Towards limb position invariant myoelectric pattern recognition using time-dependent spectral features." Neural Networks 55 (2014): 42-58.

[26] Stoica, Petre, and Randolph L. Moses. "Spectral analysis of signals." (2005).

[27] Al-Timemy, Ali H., et al. "Classification of finger movements for the dexterous hand prosthesis control with surface electromyography." IEEE journal of biomedical and health informatics 17.3 (2013): 608-618.

[28]Al-Timemy, Ali H., et al. "Improving the performance against force variation of EMG controlled multifunctional upper-limb prostheses for transradial amputees." IEEE Transactions on Neural Systems and Rehabilitation Engineering 24.6 (2015): 650-661.

[29] Too, Jingwei, Abdul Rahim Abdullah, and Norhashimah Mohd Saad. "Classification of Hand Movements based on Discrete Wavelet Transform and Enhanced Feature Extraction."(2019)

[30] Phinyomark, Angkoon, Pornchai Phukpattaranont, and Chusak Limsakul. "Fractal analysis features for weak and single-channel upper-limb EMG signals." Expert Systems with Applications 39.12 (2012): 11156-11163.

[31] Kuegler, P., S. Member, C. Jaremenko, J. Schalachelzki, 2013, "Automatic Recognition of Parkinson Disease Using Surface Electromyography During Standarized Gait Test," in IEEE International Conference on Engineering in Medicine and Biology Society (EMBC), (IEEE Explore, 2013), pp 5781-5784.

[32] Phinyomark, Angkoon, Pornchai Phukpattaranont, and Chusak Limsakul. "Feature reduction and selection for EMG signal classification." Expert systems with applications 39.8 (2012): 7420-7431.